\documentclass[aps,prl,twocolumn,
superscriptaddress,
 amsmath,amssymb
]{revtex4-1}
\usepackage{epsfig}
\usepackage{epstopdf}
\usepackage{graphicx}% Include figure files
\usepackage{dcolumn}% Align table columns on decimal point
\usepackage{bm}% bold math
\usepackage{amsmath}
\usepackage{array}
\usepackage{color}
\usepackage{titlesec}
\titleformat{\section}[hang]{\bf\scshape}{\thesection.}{1em}{}
\begin{document}

\setlength{\belowdisplayskip}{0pt} \setlength{\belowdisplayshortskip}{0pt}
\setlength{\abovedisplayskip}{0pt} \setlength{\abovedisplayshortskip}{0pt}

\renewcommand{\figurename}{{\bf Figure}}
\makeatletter 
\renewcommand{\thefigure}{{\bf \@arabic\c@figure}}
\makeatother

\title{On-chip storage of broadband photonic qubits in a cavity-protected rare-earth ensemble}
%Mapping broadband photonic qubits into on-chip rare-earth ensembles cavity-protected against decoherence} 

\author{Tian Zhong}
\affiliation{T. J. Watson Laboratory of Applied Physics, California Institute of Technology, 1200 E California Blvd, Pasadena, CA, 91125, USA}

\author{Jonathan M. Kindem}
\affiliation{T. J. Watson Laboratory of Applied Physics, California Institute of Technology, 1200 E California Blvd, Pasadena, CA, 91125, USA}

\author{Jake Rochman}
\affiliation{T. J. Watson Laboratory of Applied Physics, California Institute of Technology, 1200 E California Blvd, Pasadena, CA, 91125, USA}

\author{Andrei Faraon}
\email{faraon@caltech.edu}
\affiliation{T. J. Watson Laboratory of Applied Physics, California Institute of Technology, 1200 E California Blvd, Pasadena, CA, 91125, USA}

\date{\today}% It is always \today, today,

\maketitle

\noindent \textbf{Ensembles of solid-state optical emitters enable broadband quantum storage~\cite{Northup, Lvovsky} and transduction of photonic qubits~\cite{Williamson, Brien}, with applications in high-rate optical quantum networks for secure communications~\cite{Xie,ZhongQKD}, global time-keeping~\cite{Komar}, and interconnecting future quantum computers. To realize quantum transfer using ensembles, rephasing techniques are currently used to mitigate fast decoherence resulting from inhomogeneous broadening~\cite{Hedges,Riedmatten}. Here we use a dense ensemble of neodymium rare-earth ions strongly coupled to a nanophotonic resonator to demonstrate significant cavity protection effect at the single photon level~\cite{Kuruz, Diniz} - a new technique to suppress ensemble decoherence due to inhomogeneous broadening. The protected Rabi oscillations between the cavity field and the atomic superradiant state enable ultra-fast transfer of photonic frequency qubits into the ions ($\sim$50 GHz bandwidth), followed by retrieval with 98.7\% fidelity. With the prospect of coupling to other long-lived rare-earth spin states, this technique opens the possibilities for broadband, always-ready quantum memories and fast optical-to-microwave transducers.}

Ensembles of rare-earth ions doped in crystals exhibit outstanding quantum coherence properties and broad inhomogeneous linewidths that are suitable for quantum information transfer with fast photons \cite {Thiel}. They are used in state-of-the-art optical quantum memories with potential for microwave storage~\cite{Lvovsky, Tittel, Hedges, Riedmatten, MZhong, Wolfowicz}, and are promising candidates for optical to microwave quantum transduction~\cite{Williamson, Brien}. One major challenge towards broadband quantum interfaces based on solid-state emitters is that information stored in the collective excitation of the ensemble quickly decoheres because of inhomogeneous broadening. To restore the optical coherence, protocols based on spectral hole burning techniques like atomic frequency comb (AFC) \cite{AFC, Riedmatten, Saglamyurek} and controlled reversible inhomogeneous broadening (CRIB) \cite{Hedges} have been perfected. Although effective, these protocols involve long (hundreds of miliseconds) and complex preparation procedures that generally limit the interface bandwidth. Recently, it was proposed~\cite{Kuruz, Diniz}  that ensemble decoherence can be suppressed via strong coupling to a cavity. This phenomenon, called ‘cavity protection,’ has been experimentally observed, though not in full effect, in the microwave domain with a NV spin ensemble~\cite{Putz}. Here, we demonstrate for the first time strong cavity protection against decoherence in the optical domain using a dense ensemble (a few millions) of neodymium (Nd) atoms coupled to a nanophotonic cavity. Exploiting the protected mapping of photonic qubits to atomic superradiant excitations, we realize a 50 GHz quantum light-matter interface that could find applications in future quantum networks.

The dynamics of a coupled cavity-ensemble system are described by the Tavis-Cummings Hamiltonian~\cite{Tavis}. The interaction term reads $H_{int}=i\hbar\Omega(S^- a^{\dagger}-S^+a)$ where $a^{\dagger}$ and $a$ are creation and annihilation operators of the cavity mode, and the collective spin operators $S^{\pm}=\frac{1}{\sqrt{N}}\sum\sigma_k^{\pm}$ act on $N$ atoms each of frequency $\omega_k$. $\Omega$ denotes a collective coupling strength $\Omega^2=\sum_k^Ng_k^2$ which scales up the single atom coupling $g_k$ by $\sqrt{N}$. On resonance, the coupled system exhibits two bright polariton states with equal mix of atomic and photonic components detuned by $\pm \Omega$ from the mean ensemble frequency. The polaritons decay via radiative emission and decohere by coupling to dark subradiant states that overlap spectrally with the ensemble~\cite{Houdre, Kuruz, Diniz}. The dark-state coupling critically depends on the energy separation between the polaritons and the subradiant states, and also on the specific profile of the inhomogeneous spectral distribution $\rho(\omega)=\sum_k g_k^2\delta(\omega-\omega_k)/\Omega^2$ \cite{Kuruz, Diniz}. In the limiting case of a Lorentzian distribution, considerable damping given by the width of the inhomogeneous broadening persists even with an infinite $\Omega$. When the spectral distribution exhibits a faster-than-Lorentzian decay (e.g. Gaussian), the damping of the coherent Rabi oscillation is diminished at increasing $\Omega$ - the system becomes `cavity protected' as conceptually illustrated in Fig.~1a. In this case, the atomic component of the polariton is purely the symmetric superradiant state~\cite{Dicke}.

\begin{figure*}[htb]
\includegraphics[width=.95\textwidth]{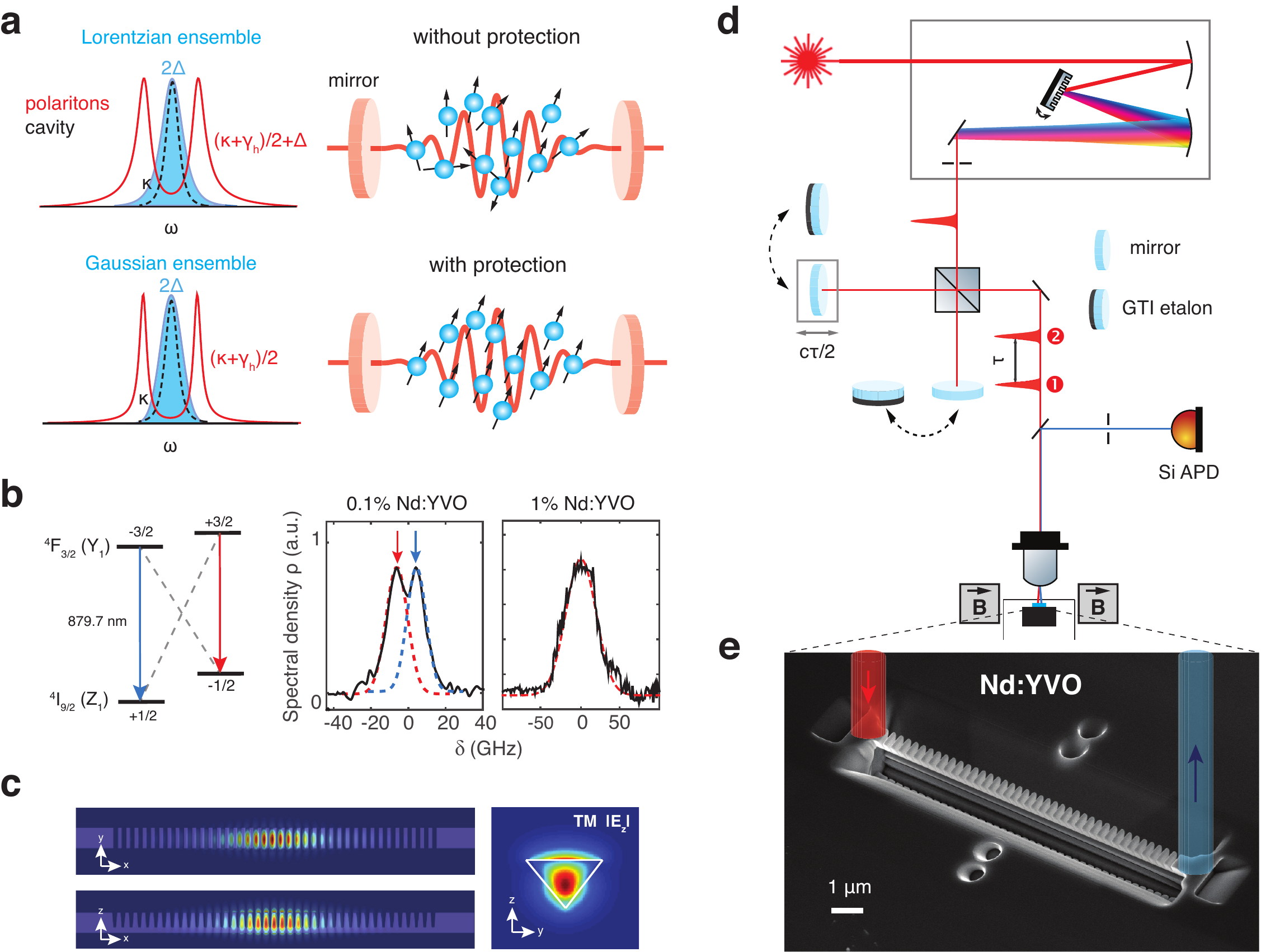}
\caption{{\bf Schematics of the cavity protection effect and its experimental setup.} {\bf a,} Conceptual illustration of cavity protection for an ensemble coupled to a cavity mode. For a Lorentzian ensemble (upper), the polaritons are not protected and undergo dephasing (linewidth broadening) due to inhomogeneous broadening $\Delta$. A Gaussian ensemble (lower) can be fully protected with the collective superradiant excitation free of such dephasing. Arrows represent the phasor of each atomic dipole. {\bf b,} Energy levels and transitions (dotted lines are forbidden) for Nd (left). Measured absorption spectra for 0.1\% and 1\% Nd:YVO (right). Two Zeeman split sub-ensembles are resolved in the 0.1\% sample. {\bf c,} Simulated TM resonance mode profiles of the triangular nanobeam resonator. {\bf d,} Experimental setup. Two pico-second pulses were transmitted through the cavity and the output signal was integrated on a Si APD photon counter. (e) Scanning electron microscope image of the device and schematics of input and output optical coupling.}\label{f1}
\end{figure*}

We probe the cavity-protection regime in an optical nanocavities based on our triangular beam design \cite{Zhongfab, Zhong} fabricated in Nd-doped yttrium vanadate (YVO) crystals (Fig.~1e). The cavity has a fundamental TM mode resonance with measured quality factor Q of 7,700 ($\kappa\sim2\pi\times$44 GHz is the energy decay rate, full-width at half-maximum (FWHM) is 44 GHz) and 17,000 ($\kappa\sim2\pi\times$20 GHz, FWHM=20 GHz) in 0.1\% and 1\% Nd:YVO, respectively. A simulated mode volume $V_{\rm mode}=1(\lambda/n)^3$=0.063 $\mu$m$^3$ estimates $N\sim10^6$($10^7$) ions in the 0.1\%(1\%) cavity. The resonance wavelengths are close to the $^4I_{9/2} (Y_1)-^4F_{3/2}(Z_1)$ transition of Nd$^{3+}$ at 879.7 nm. The devices were cooled down to 3.6 K (Montana Instruments Cryostation) and a magnetic field of 500 mT was applied perpendicular to the YVO c-axis. In 0.1\% Nd:YVO, the optical coherence time is T$_2$=390 ns (measured via photon echoes), corresponding to a single emitter homogeneous linewidth $\gamma_h/2\pi=1/\pi T_2=$0.82 MHz. In 1\% Nd:YVO, we measured an upper bound of $\gamma_h/2\pi\le40$ MHz via transient hole burning. The {\bf B} field caused a Zeeman splitting of the $Y_1, Z_1$ states into 4 levels (Fig.1b). For our field orientation, cross-transition probabilities are minimized~\cite{AfzeliusYVO}. Therefore the system can be viewed as two independent distributions of emitters separated by 17 GHz (shown as resolved absorption lines in 0.1\% but not in 1\% device) both coupled to the cavity with similar strengths. To capture the spectral shape of the distribution, a $q$-Gaussian function was used to fit each transition~\cite{Putz}, yielding a shape parameter $q$=1.01 (1 for Gaussian, 2 for Lorentzian) for the 0.1\% ensemble. Each Zeeman branch has a FWHM of $\gamma_q=2\Delta\sqrt{\frac{2^q-2}{2q-2}}=2\pi\times$8.3 GHz with $\Delta/2\pi$=5.0 GHz ($\Delta/\pi$ represents the FWHM for a Lorentzian distribution). The total FWHM of the ensemble including both branches is 24 GHz. The 1\% ensemble exhibits an asymmetric distribution with 76 GHz FWHM. However, it cannot be fitted well with any common functions because at this concentration the ions exhibit various interactions between themselves and with crystalline defects that lead to satellite lines.
 
To achieve strong coupling, the cavity resonance was tuned towards longer wavelengths by gas condensation~\cite{Zhong} while the transmission from a broadband superluminescent input was recorded in Fig.~2a,d using a spectrometer. The on-resonance spectra (Fig.~2b,e) reveal two bright polariton peaks with a Rabi splitting of $\Omega_R/2\pi=$110 GHz and 48 GHz, for the 1\% and 0.1\% device, respectively. In Fig.~2e, a middle peak is present in between the polaritons because the cavity coupled simultaneously to two Zeeman branches with resolved splitting (see Supplementary Information II). The decay rates $\Gamma(\delta)$ were determined from the FWHM linewidth of the left polariton peak, and are plotted against the cavity-ensemble detuning $\delta=\omega_c-\omega_a$ in Fig.~2c,f as black triangles. The data corresponds to the left anti-crossing trajectory in Fig.~2a,d as the cavity shifted from shorter wavelengths towards the atomic resonance. 

The phenomenon of cavity protection can be observed in both concentration samples, as the on-resonance $\Gamma$ is considerably narrower than the Lorentzian (no protection) limit $\kappa/2+\gamma_h/2+\Delta$. In the 1\% sample, $\Gamma(0)/2\pi=$21 GHz is 35 GHz narrower than the Lorentzian limit and also narrower than the FWHM of the initial inhomogeneous broadening of the ensemble, thus indicating that the polariton decay is slower than the decoherence of the initial ensemble. We point that in Fig.~2c (1\% Nd:YVO), $\Gamma/2\pi$ slightly increases around $\delta=$50 GHz before decreasing to a minimum of 21 GHz on resonance. That increase might be explained by coupling to one of the Nd-Nd pair site that is known to be blue-detuned from the central transition by 48 GHz~\cite{DM} (Supplementary III). The data for the 1\% sample is not compared with a theoretical model because, as mentioned above, the exact distribution of ions is unknown. For the 0.1\% sample shown in Fig.~2f, $\Gamma(0)/2\pi=$22 GHz and the data shows good agreement with the theoretical decay (red curve) for a Gaussian-distributed ensemble with the same FWHM as the joint distribution of the two Zeeman branches (Supplementary Information I, II). In the on resonance, strong coupling limit, the theoretical decay is expressed as $\Gamma=\kappa/2+\gamma_h+\pi\Omega^2 \rho(\Omega)$~\cite{Diniz}, which reaches the full protection limit of $\Gamma=\kappa/2+\gamma_h=2\pi\times$22 GHz as indicated in Fig.~2f. In our case, the experimental data approached this limit. The residual broadening estimated from the $\pi\Omega^2 \rho(\Omega)$ term was $\approx$0.1 GHz, more than two orders of magnitude suppressed than without protection (the residual broadening would be 14.6 GHz). Although close to fully protected, the total decay rate was not much slower than the initial ensemble decoherence (FWHM 24 GHz ($\Delta/2\pi$=14.6 GHz) by treating the two Zeeman branches as one joint distribution). To contrast with no protection scenarios, we also plot in green the theoretical decays of upper and lower polaritons for a Lorentzian distribution (Supplementary Information I) assuming the same $\Delta$ as for our ensembles, of which the atom- and cavity-like polariton widths converge to the Lorentzian limit at zero detuning.

\begin{figure*}
\includegraphics[width=.95\textwidth]{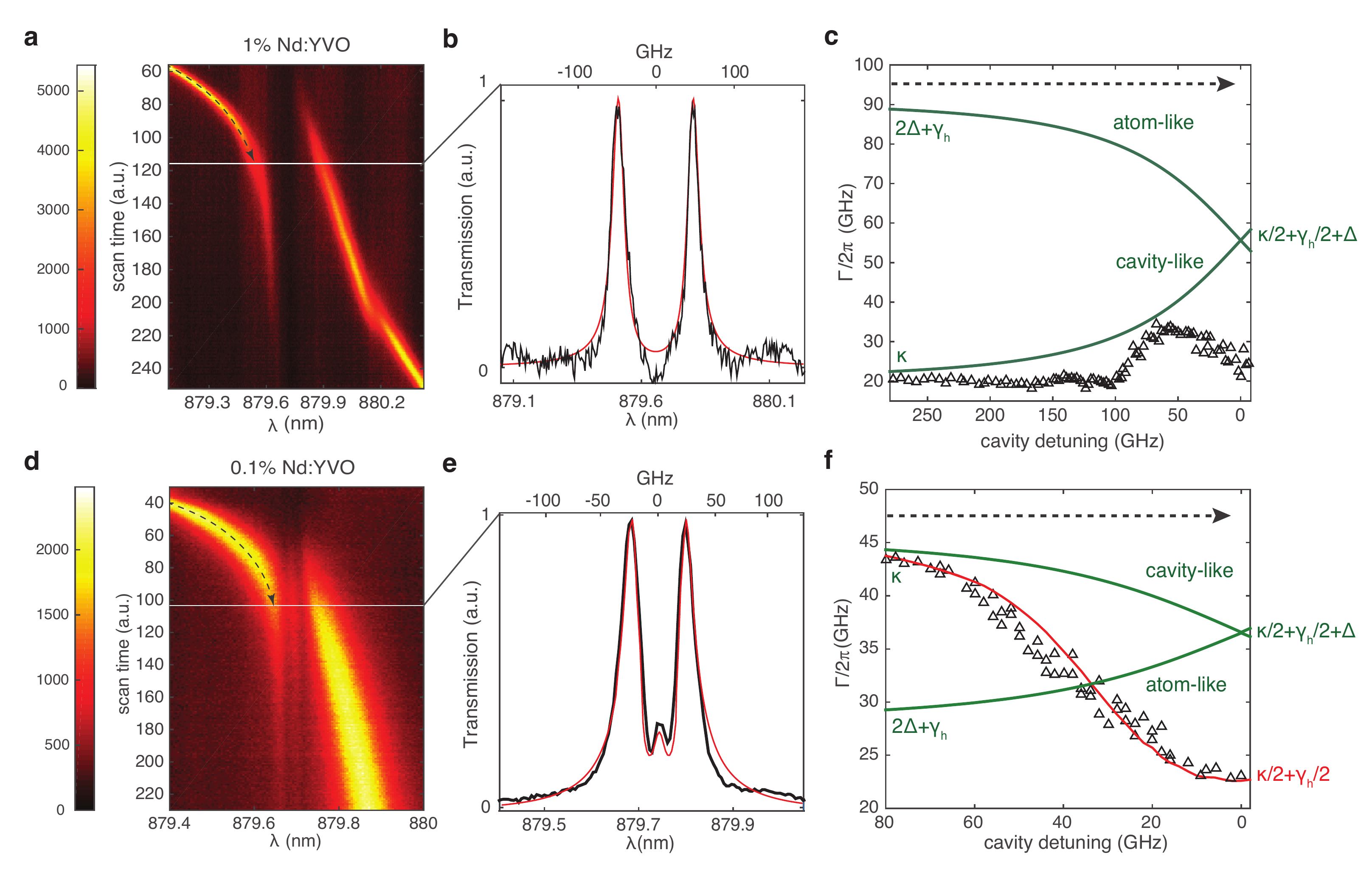}
\caption{{\bf Cavity protection of the Nd ensemble against decoherence.} {\bf a, d} Cavity transmission spectra while tuning its resonance across the inhomogeneous Nd transition. {\bf b, e} On-resonance transmission spectra showing two bright polaritons. Red curves are theoretical fit assuming Gaussian ensembles. {\bf c, f} Experimental decay rates extracted from the left anti-crossing trajectory (dotted arrow) in {\bf a} as a function of detuning. In {\bf c} (1\% Nd:YVO), the polariton (21 GHz) is significantly narrower than the initial inhomogneous broadening (76 GHz), but it is not reaching the full protection limit likely due to an asymmetric, non-Gaussian ensemble shape. In {\bf f} (0.1\%), polariton linewidth decreases rapidly towards resonance to the full protection limit ($\kappa/2+\gamma_h/2$). Red curve plots the theoretial decay for a Gaussian distribution. Green curves show the theoretical decays assuming a Lorentzian ensemble of the same $\Delta$.}\label{f2}
\end{figure*}

The cavity-protected system acts as a quantum interface where a broadband photon can be transferred to the superradiant atomic excitation. We measured this coherent, ultra-fast dynamics using pulsed excitations of the polaritons. The experimental setup is depicted in Fig.~1d. A mode-locked Ti:Sapphire laser at 85 MHz repetition rate (Thorlabs Octavius)  was filtered to a pulse width of 4(1.5) ps using a monochromator, which was sufficient to simultaneously excite both upper ($| \omega_+ \rangle \rightarrow |1\rangle$) and lower ($| \omega_- \rangle \rightarrow  |0 \rangle$) polaritions in 0.1\% (1\%) device. The filtered laser was attenuated and sent through a Michelson setup to produce two pulses with less-than-one mean photon number separated by a variable delay $\tau$ that were coupled into the cavity (red path) and the transmitted signal was collected (blue path) for direct detection using a silicon single photon counter. The integrated counts at varying delays produces optical field autocorrelation signals revealing the temporal evolution of the polaritons. The mirror at each Michelson arm was interchangeable with a Gires-Tournois Interferometer (GTI) etalon, which generates a $\sim\pi$/2 phase chirp between the two polaritons (Methods). Furthermore, a narrow bandpass filter was optionally inserted in either arm that allowed only one polariton to be excited. This combination enabled a comprehensive polariton excitation scheme that covered individual polariton $|0\rangle$ or $|1\rangle$, and superposition states of two polaritons i.e. $|+\rangle=1/\sqrt{2}(|0\rangle+|1\rangle)$ or $|\circlearrowleft\rangle=1/\sqrt{2}(|0\rangle+i|1\rangle)$. 

\begin{figure*}[htb]
\includegraphics[width=.85\textwidth]{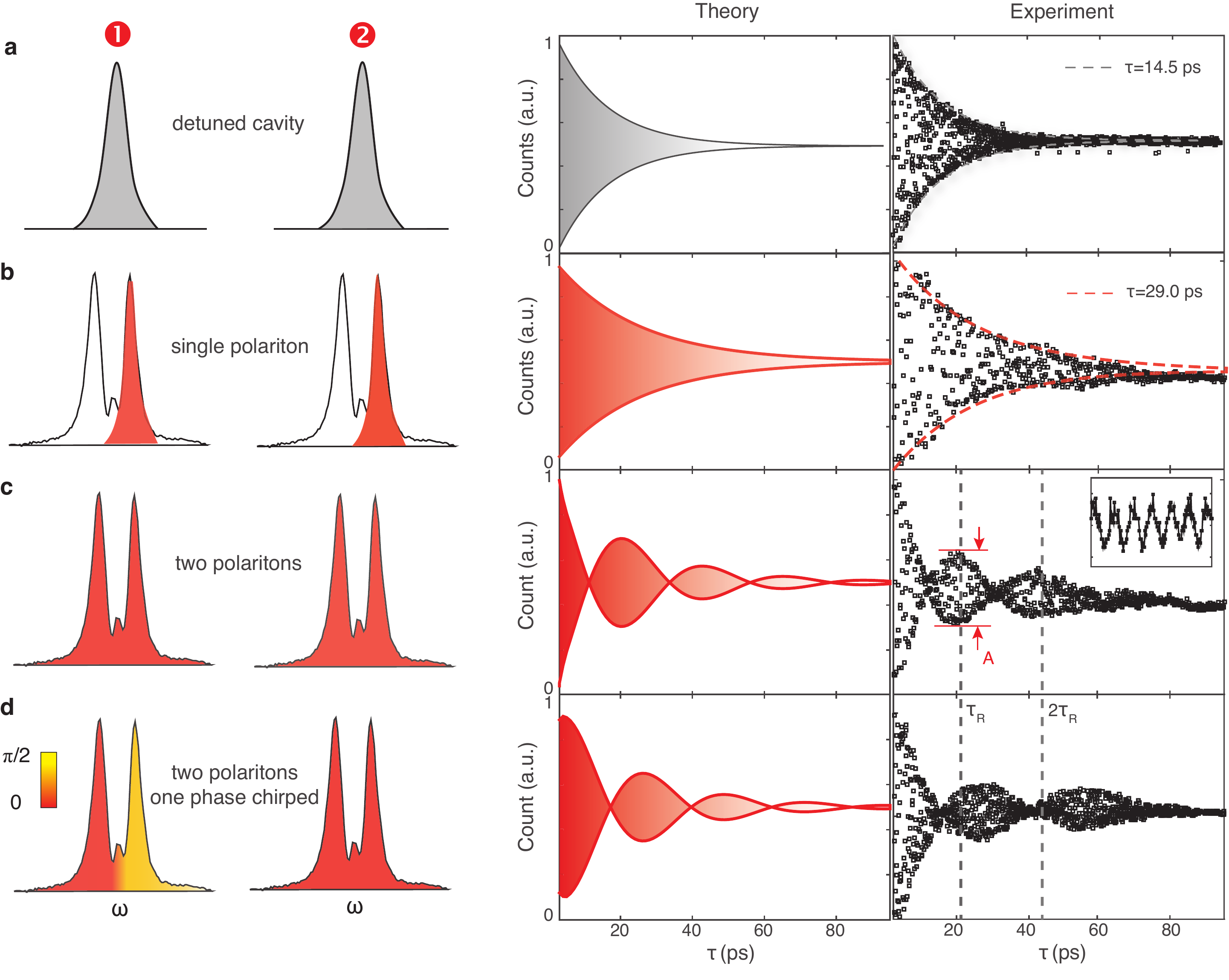}
\caption{{\bf Time-domain interferometric measurements of the cavity transmission.} Left panels show the cavity transmission spectra (black outlines) of the probed system. Colored areas show the spectral range addressed by the probe pulses. Grey is for uncoupled cavity; red for excited polariton states: yellow for the polarition with a shifted phase. {\bf a,} Simulated and measured cavity decay (i.e. lifetime) when uncoupled from the atoms. {\bf b,} Decay of singly excited lower (upper) polarition (i.e. $|0\rangle(|1\rangle)$). {\bf c,} Decayed oscillations when both polaritons were excited with transform-limited pulses. {\bf d,} Decayed and time-shifted oscillations when two polaritons were initially excited with a $\pi$/2 phase difference (i.e. $|0\rangle+i|1\rangle$) by the first pulse and in phase by the second pulse. {\bf b-d} shows extended decays about twice in {\bf a}, confirming a nearly full protection against ensemble-induced decoherence. The dotted lines mark multiples of Rabi periods $\tau_R$. The inset shows a few fine fringes scaned around $\tau_R$. }\label{f3}
\end{figure*}

Figure 3 plots the theoretical interference fringe amplitudes along with the measured results for several two-pulse excitation schemes for the 0.1\% cavity in which maximum protection was observed. The results for the 1\% device are presented in Supplementary Information VI. The mean photon number per pulse coupled in the cavity was estimated at $\mu$=0.5. The case of an uncoupled cavity is plotted in Fig.~3a, showing a fitted decay constant ($4/\kappa$) of $\sim$14.5 ps (Supplementary Information V). When only one polariton was excited, the decay was extended to 29.0 ps (Fig.~3b). For the superposition state $|+\rangle$, Ramsey-like fringes were obtained, revealing extended Rabi oscillations between photonic and atomic excitations beyond the cavity lifetime (Fig.~3c). In the case of Fig.~3d, the first pulse excited the two polaritons with a phase chirp. The resulting fringe showed the Rabi oscillations with the nodes shifted with respect to 3c by about 5.5 ps ($\sim1/4\tau_R$), in agreement with our theoretical model. Those nodes correspond to the quantum excitation being entirely stored in the ensemble with no energy left in the cavity mode, during which time the stored qubit dephases at a significantly slower rate than the inhomogeneous broadening. 

\begin{figure*}[htb]
\includegraphics[width=0.85\textwidth]{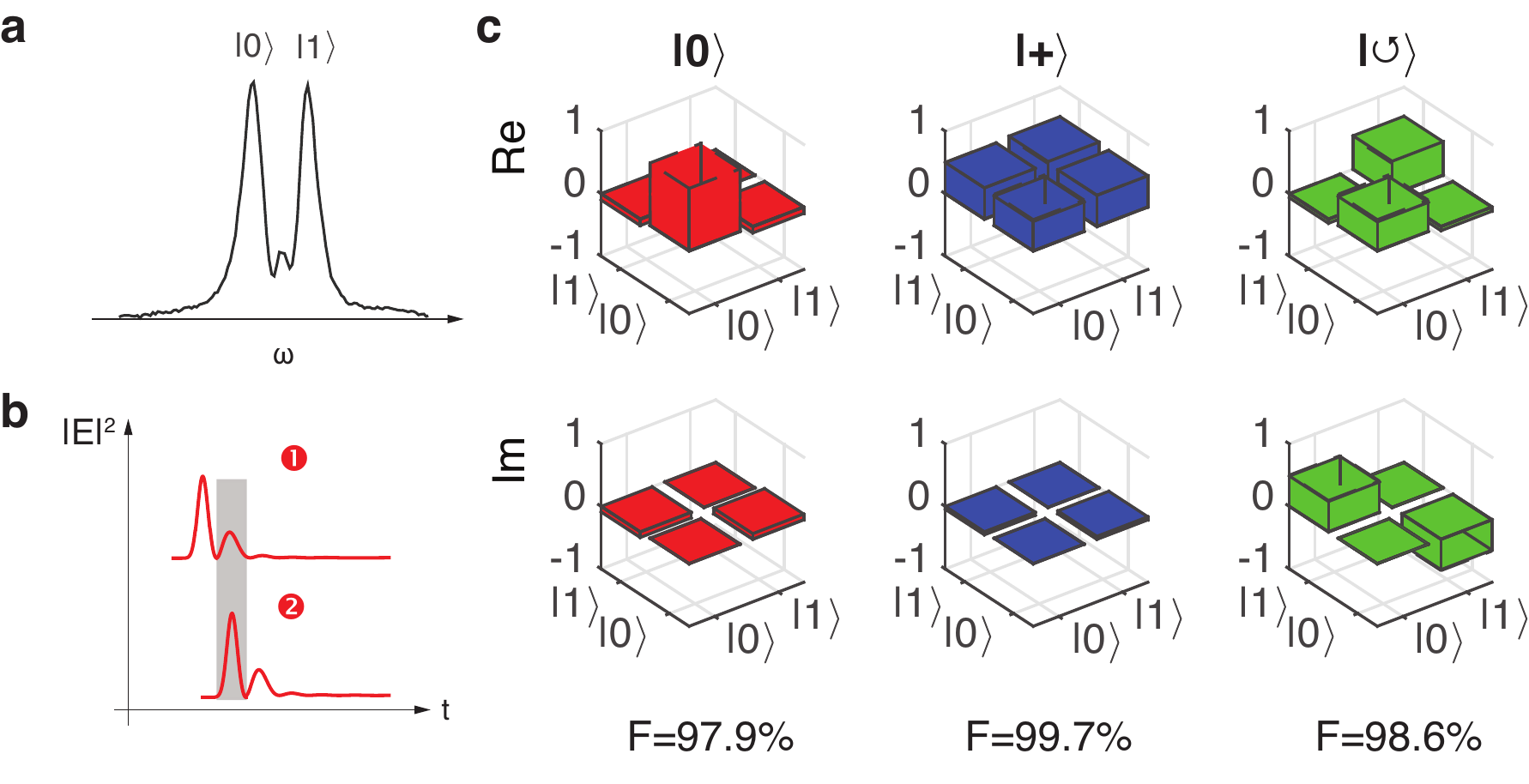}
\caption{{\bf Broadband qubit transfer to the protected Nd ensemble.} {\bf a,} Two polaritons serve as eigen basis for a frequency bin qubit. A phase chirp can be added to construct a generic qubit $|0\rangle+e^{i\phi}|1\rangle$. {\bf b,} Simulated time-domain evolution of the cavity field intensities of two qubit-encoded photons, showing temporal overlap between the retrieved photon ($|\psi_{1}\rangle$) and the second photon ($|\psi_{2}\rangle$) directly transmitted through the cavity at time $\tau_R$. The fields dominantly overlap within a temporal window (grey area) when interference occurs yielding integrated photon counts proportional to the overlap $|\langle\psi_{2}|\psi_{1}\rangle|$. The overlap after the grey window was negligibly small due to fast cavity decays. {\bf c,} Reconstructed density matrices for each input qubit (top row) delayed by $\tau_R$ showing high fidelities.}\label{f4}
\end{figure*}

This quantum interface is similar to an AFC with two teeth, one at each polariton, that form the basis of a frequency bin qubit as shown in Fig.~4a. Photons are stored and then released after inverse of the teeth spacing, which is a Rabi period $\tau_R$. The interface bandwidth is $\sim$50 GHz, spanning two polaritons, and the qubits are of the form $|0\rangle$, $|+\rangle=1/\sqrt{2}(|0\rangle+|1\rangle)$ or $|\circlearrowleft\rangle=1/\sqrt{2}(|0\rangle+i|1\rangle)$.  To characterize this process, quantum state tomography on the released qubit after a delay $\tau_R$ was performed. As direct projection measurements were difficult given the high-bandwidth, we adopted an interferometric scheme (Fig.~4b) to assess the input/output fidelity $F=\langle\psi_{in}|\rho_{out}|\psi_{in}\rangle$, where $|\psi_{in}\rangle$ is the input qubit state and $\rho_{out}$ is the density matrix for the retrieved state, from a set of fringe signals including those in Fig.~3 (Methods and Supplementary Information VII). The reconstructed density matrices $\rho_{out}$ for $|0\rangle$, $|+\rangle$, and $|\circlearrowleft\rangle$ input states along with their respective fidelities are shown in Fig.~4c. A mean fidelity of 98.7$\pm$0.3\% is obtained, which significantly surpasses the classical fidelity limit (the best qubit input/output fidelity one can achieve using a classical intersect-resend strategy~\cite{Massar}) of 74.9$\pm$0.04\% that takes into account the Poissonian statistics of the coherent input photons (with $\mu$=0.5) and an imperfect but high storage-retrieval efficiency of 25.6$\pm$1.2\% (Methods)~\cite{Specht, Gundogan}. The estimated fidelities take into account imperfections in the preparation and measurement of the qubit, such as leakage of traveling waves through the cavity and inaccurate phase shift (ideally $\pi$/2) by the GTI etalon. Thus, the high fidelity indicates a robust quantum transfer with a bandwidth that is significantly broader than existing rare-earth-based light-matter interfaces~\cite{Tittel, Saglamyurek}. %To highlight the benefit of cavity protection, we also evaluated the qubit fidelity at a delay of 2$\tau_R$, which would be equivalent to the case without cavity protection where the qubit would decohere twice as fast. The measured fidelities at 2$\tau_R$ dropped to 83\% for $|0\rangle$, 70\% for $|+\rangle$, and 69\% for $|\circlearrowleft\rangle$, which no longer beats the classical fidelity. 

While this interface efficiently maps the photonic qubit to the ensemble, the qubit dissipates at $\kappa/2$ rate. Improvements in the cavity quality factor to state of the art values of Q$\sim10^6$ would achieve storage for 1 ns (enough for perform 50 Rabi flips). To enable long-term storage like in an AFC-spin-wave memory~\cite{AFC}, the qubit can be transferred from the superradiant state excitation to a long-lived spin level by applying a $\pi$ pulse within $\tau_R$ time.  Upon recall, another $\pi$ pulse can transfer the qubit back to the polariton states and then a cavity photon. For faithful spin-wave storage, the Rabi frequency of the $\pi$ pulses should exceed the polariton linewidths, which is attainable given the strong light confinement in current nanobeam devices. Compared to existing AFC-spin-wave memories, this interface would not require any preparation steps or time-multiplexing to achieve always-ready operation. Taking advantage of on-chip platforms also allow spatially and temporally multiplexed storage by routing photons to an array of nanocavities with different delays. Most notably, the cavity-protected mapping of a photonic qubit to a collective superradiant state could compliment the reported coupling of rare-earths to a superconducting resonator \cite{Bushev} to fulfill efficient quantum transduction between optical and microwave photons via Zeeman or hyperfine transitions in rare-earth ensembles~\cite{Brien, Williamson}.

\newpage
\section*{Methods}
\noindent {\bf Nanocavity design and characterization.} The triangular nanobeam has a width of 770 nm and length of 15 $\mu$m. 40 periodic subwavelength grooves of 185 nm along the beam axis were milled on top of the nanobeam. The period of the grooves were modulated at the center of the beam to form defect modes in the photonic bandgap. The fundamental TM mode, with side, top and cross-section views shown in Fig.~1c, is chosen because it aligns with the strongest dipole of the 879.8 nm transition in Nd:YVO. The theoretical quality factor is 300,000 with a mode volume of 1($\lambda$/n)$^3$~\cite{Zhongfab}. Transmission of the nanocavity was measured by vertically coupling free-space input into the nanobeam via a 50x objective lens and a 45$^{\circ}$-angled reflector milled into the sample surface, and the cavity output was collected via the other reflector which sent the light back vertically to free space. The output signal was effectively isolated from the input reflections or other spurious light by spatial filtering using a pin hole. When the cavity is on resonance, we measured a total transmission (from free-space input to output) of 20\%, which was primarily limited by the imperfect coupling into the nanobeam. The output signal also contained leakage travelling waves (5\%) due to finite extinction of the photonic bandgap and other spurious reflections in the system.

%setup, Michelson, GT etalon, bandpass filter
\noindent {\bf Polariton excitation and frequency qubit preparation.} The GTI etalon was made of a 250 $\mu$m thick quartz slide with backside coated with a gold film. The front side was uncoated, which has a reflectivity of 4\%. This etalon produces a nearly linear dispersion of 4$\pi$/nm over a free spectral range of 0.5 nm near 880 nm. After the GTI etalon, the transform-limited laser pulse acquired a phase chirp, which excited a mixed polariton state approximated by  $|0\rangle+e^{i\phi}|1\rangle$, where $\phi$ is the phase shift over the Rabi splitting. For our custom made etalon, $\phi\approx0.52\pi$ and the corresponding polarition state was close to $(|0\rangle+i|1\rangle)$. 

%QST
\noindent {\bf Quantum state tomography based on two-pulse interferometry.} The electric field operators for the two consecutive photonic states $|\psi_1\rangle, |\psi_2\rangle$ coupled to the cavity are written as $\hat{E}_1(t)=\alpha_1 e^{-i\omega_-t}\hat{a_-}+\beta_1 e^{-i\omega_+t}\hat{a_+}$ and $\hat{E}_2(t)=\alpha_2 e^{-i\omega_-t}\hat{a_-}+\beta_2 e^{-i\omega_+t}\hat{a_+}$, respectively. The field operator at the single photon detector is $\hat{E}(t)=\hat{E}_1(t-\tau_R)+\hat{E}_2(t)$ corresponding to the first photon delayed by $\tau_R$ after storage in the ensemble, which interferes with the second photon. The count rate on the detector is $C=\langle E(t)^{\dagger} E(t)\rangle=2+2{\rm cos(\phi)}|\alpha_1\alpha_2^*+\beta_1\beta_2^*|$, where $\phi$ is the phase difference between the two photon wavepackets. This gives interference fringes with an amplitude $A=C_0|\alpha_1\alpha_2^*+\beta_1\beta_2^*|$ as labelled in Fig.~3c, where $C_0$ is a constant factor. By encoding a set of four basis states (Pauli tomography basis) on the second photon, i.e. $\alpha_2|0\rangle+\beta_2|1\rangle$,  we construct the set of experimental amplitude parameters $A_j, j=0...3$ ($A_0$ for $|\psi_2\rangle=1/\sqrt{2}|\psi_1\rangle$; $A_1$ for $|\psi_2\rangle=|+\rangle$, $A_2$ for $|\psi_2\rangle=|\circlearrowleft\rangle$; $A_3$ for $\psi_2\rangle=|0\rangle$) which are analogous to the set of projection measurement outcomes for calculating the density matrix $\hat{\rho}_{out}$ (see Supplementary Information VII for detailed derivations)
\begin{equation}
\hat{\rho}_{out}=\frac{1}{2}[\hat{I}+((A_1/A_0)^2-1)\hat{\sigma}_1+((A_2/A_0)^2-1)\hat{\sigma}_2+((A_3/A_0)^2-1)\hat{\sigma}_3.]
\end{equation}
where $\hat{I}$ is the identity operator and $\hat{\sigma}_{1,2,3}$ are the Pauli spin operators. Then we perform a maximal likelyhood estimation \cite{James}  to obtain a physical density matrix, which is used to calculate the fidelity $F=\langle\psi_{in}|\hat{\rho}_{out}|\psi_{in}\rangle$

\noindent {\bf Qubit storage and retrieval efficiency.} The storage efficiency is defined as the probability that a photon in the cavity mode is transferred to a collective excitation in the ensemble. This storage efficiency is intrinsically 100\% as the polariton modes under cavity protection condition are maximally entangled states between the cavity and the superradiant state. The retrieval efficiency is defined by the number of photons emitted at the cavity output during the second Rabi period (grey window in Fig.~4b) vesus the total transmitted photons. Based on the temporal distribution of the transmitted photons (deconvolved from the oscillation signal in Fig.~3c), the integrated counts during the second Rabi period (grey window in Fig.~4b) was 25.6$\pm$1.2\% of the total transmitted photons. Note that this storage and retrieval efficiency does not take into account the input coupling and scattering loss. Including the 20$\pm$2\% transmission efficiency through the device, the overall system efficiency was 5.1$\pm$0.7\%. The corresponding classical bound for qubit storage fidelity would be 78.9$\pm$0.05\%, still significantly below the measured fidelities in Fig.~4c.

\section*{Acknowledgements}
This work is funded by NSF CAREER 1454607, NSF Institute for Quantum Information and Matter PHY-1125565 with support from Gordon and Betty Moore Foundation GBMF-12500028, and AFOSR Quantum Transduction MURI FA9550-15-1-002. Device fabrication was performed in the Kavli Nanoscience Institute with support from Gordon and Betty Moore Foundation.

\section*{Author contributions}
A.F. and T.Z. conceived the experiments. T.Z. and J. R. fabricated the device. T.Z. and J.M.K. performed the measurements and analyzed the data. T.Z. and A.F. wrote the manuscript with input from all authors.

\section*{Competing financial interests} The authors declare no competing financial interests.

\clearpage

\newcommand{\beginsupplement}{%
        \setcounter{table}{0}
        \renewcommand{\thetable}{S\arabic{table}}%
        \setcounter{figure}{0}
        \renewcommand{\thefigure}{S\arabic{figure}}%
     }

\onecolumngrid

      \beginsupplement

\section*{On-chip storage of broadband photonic qubits in a cavity-protected rare-earth ensemble: Supplementary Information}

\subsection{Theoretical decay rates of coupled cavity-ensemble system with detuning}
\noindent The cavity-ensemble coupled system is described by the Tavis-Cummings Hamiltonian. We model the system following \cite{Diniz}, which consists of a cavity mode $a$ of frequency $\omega_0$ coupled with strength $g_k$ to a distribution of $N$ two-level emitters described by modes $b_k$. We define the cavity frequency $\omega_0$ and the frequency of each emitter as $\omega_k$.  We account for the atomic dephasing rate (homogeneous linewidth) $\gamma_h$ and label the cavity intensity decay as $\kappa$. Using the standard input-output formalism for a two sided cavity with input field $c_{in}$, reflected field $c_r$, and transmitted field $c_t$ gives the Heisenberg equations for the system:
\begin{eqnarray}
\dot{a} &=& -\left[\frac{\kappa}{2} + i (\omega_0 - \omega)\right] a - \sqrt{\frac{\kappa}{2}}c_{in} + \sum_k g_k b_k\nonumber\\
\dot{b}_k &=& -\left[\frac{\gamma_h}{2} + i (\omega_k - \omega)\right] b_k - g_k a \nonumber\\
c_t &=& \sqrt{\frac{\kappa}{2}}a\nonumber\\
c_r &=& c_{in} + \sqrt{\frac{\kappa}{2}}a.
\end{eqnarray}

\noindent By solving this set of equations in the steady state, we arrive at the complex transmission of the cavity:
\begin{equation}
t(\omega) = \frac{\left<c_t\right>}{\left<c_{in}\right>}=\frac{- \frac{\kappa}{2 i}}{\omega_0 - \frac{ i \kappa}{2} - \omega - \sum_k \frac{g^2_k}{\omega_k - \frac{i \gamma_h}{2} - \omega}}.
\end{equation}
For a large number of emitters, we can define the distribution of emitters in terms of a continuous spectral density $\rho(\omega) = \sum_k \frac{g_k^2 \delta(\omega - \omega_k)}{\Omega^2}.$  Here $\Omega$ is the collective coupling strength defined by $\Omega^2 = \sum_k g_k^2 $. The continuum form of the transmission is then \cite{Diniz}

\begin{equation}
t(\omega) = \frac{- \frac{\kappa}{2 i}}{\omega_0 - \frac{ i \kappa}{2} - \omega +\Omega^2 \int \frac{\rho(\omega')d\omega'}{\omega - \omega' + \frac{i\gamma_h}{2}}}.
\end{equation}

\noindent {\bf Lorentzian distribution} In the case of a Lorentzian distribution, $\rho(\omega)=\frac{\Delta}{\pi}\frac{1}{\Delta^2+(\omega-\omega_a)^2}$, where $2\Delta$ is the FWHM of the inhomogeneous linewidth and $\omega_a$ denotes the center frequency of the ensemble, Eq.~(3) is integrable to 

\begin{equation}
t(\omega)_{\rm Lorentzian} = \frac{- \frac{\kappa}{2 i}}{\omega_0 - \frac{ i \kappa}{2} - \omega + \frac{\Omega^2}{\omega - \omega_a +i\gamma_h/2 + i\Delta}}.
\end{equation}

\noindent The poles of the transmission function yield
\begin{equation}
\omega_\pm = \omega_a+\delta/2 - i \frac{\kappa+\gamma_h + 2\Delta}{4} \pm \sqrt{\Omega^2+[(i\kappa-2i\Delta-i\gamma_h-2\delta)/4]^2}
\end{equation}

\noindent where we define the cavity-ensemble detuning as $\delta=\omega_0-\omega_a$. Eq.~(5) determines the locations and linewidths of the two polariton modes: 
\begin{equation}
\Gamma_{\pm}=\frac{\kappa+\gamma_h+2\Delta}{2} \pm {\rm Im}(\sqrt{4\Omega^2+[(i\kappa-2i\Delta-i\gamma_h-2\delta)/2]^2})
\end{equation}

\noindent which checks $\Gamma=\frac{\kappa+\gamma_h+2\Delta}{2}$ for the on-resonance ($\delta=$0), strong coupling ($\Omega\gg\kappa,\Delta$) condition. The green curves in Fig.~2{\bf c},2{\bf f} are thereby evaluated from Eq.~(6).

\noindent {\bf Gaussian distribution} In the case of a Gaussian distribution, $\rho(\omega)=\frac{1}{\Delta \sqrt{\pi}}e^{-(\omega-\omega_a)^2/\Delta ^2}$. Note that the definition of $\Delta$ here is different from that in \cite{Diniz}. We define $\Delta$ such that the FWHM of the Gaussian distribution is $2\sqrt{\rm ln 2}\Delta$, which is consistent with the q-Gaussian definition in the text. Integrating Eq.~(2) gives
\begin{equation}
t(\omega, \delta)_{\rm Gaussian} = \frac{- \frac{\kappa}{2 i}}{\omega_a + \delta - \frac{ i \kappa}{2} - \omega - i\frac{\Omega^2}{\Delta}\sqrt{\pi}e^{-(\frac{\omega-\omega_a+i\gamma_h/2}{\Delta})^2}\rm erfc(-i\frac{\omega-\omega_a+i\gamma_h/2}{\Delta})}.
\end{equation}

\noindent where erfc is the complex complementary error function. There is no straightforward analytical solution for the poles of the transmission function. Therefore we numerically solve for the transmission poles and the polariton linewidth as a function of the cavity-ensemble detuning $\delta$.

To plot the theoretical decay rates in Fig.~2{\bf c},2{\bf f}, we need experimental values for $\kappa$, $\Delta$, $\Omega$ and $\gamma_h$. In the case of a far detuned cavity, using a spectrometer (with resolution of 4 GHz) we directly measured  $\kappa$ = 2$\pi\times$44 GHz for the 0.1\% Nd:YVO device, and $\kappa$ = 2$\pi\times$20 GHz for the 1\% Nd:YVO device. As discussed in the main text, the atomic distributions are approximately Gaussian, with $\Delta/2\pi=$14.6 GHz from fitting the two branches with one Gaussian for the 0.1\% Nd:YVO device, and a $\Delta/2\pi$=45.6 GHz for the 1\% Nd:YVO device. We then measured $\gamma_h/2\pi$=1/$\pi$T$_2$= 0.82 MHz for 0.1\% Nd:YVO using two-pulse photon echoes. We could not measure any echo signal from the 1\% Nd:YVO device, probably due to significant ion-ion interactions in a highly doped sample. However, we estimated an upper bound of $\gamma_h/2\pi\le40$MHz by transient hole burning in 1\% sample. To determine $\Omega$, we fit the on-resonance transmission spectra (Fig.~2{\bf b,e}) with $\Omega$ as the only free parameter. Fitting based on Eq.~(7) allows us to determine the collective coupling strength $\Omega$ = $2\pi\times$25 GHz for the 0.1\% device, and $\Omega$ = $2\pi\times$55 GHz for the 1\% device. 

\subsection{Observation and explanation of the middle peak in 0.1\% Nd:YVO nanocavity}
\noindent When tuning the cavity across the Nd Y$_1$-Z$_1$ transition in the 0.1\% device, we observed a weak transmission peak between the two polariton modes. This peak is caused by coupling a cavity to two distinct ensembles of emitters spectrally separated by 2$\delta_a$, which was theoretically predicted and analysed in Section VI. B of \cite{Diniz}. This middle peak corresponds to the eigenstate resulting from the coupling between the cavity mode and the antisymmetric state $|A\rangle=(|G_1,S_2\rangle-|S_1,G_2\rangle)/\sqrt{2}$, where $G_j, S_j$ are ground and excited state of the superadiance collective state of $j$th ensemble. According to \cite{Diniz}, the state giving rise to the middle peak is written as $i\delta_a|1,G_1,G2\rangle+\Omega\sqrt{2}|0,A\rangle$. This state would be completely dark if $\Omega\gg\delta_a$, and has a small cavity component otherwise, which we show below is the case in our 0.1\% Nd:YVO device.

To model the system, we start with a atomic distribution as a summation of two Gaussian distributions spectrally separated by 2$\delta_a$: $\rho(\omega)=\frac{1}{2\Delta \sqrt{\pi}}e^{-(\omega-\omega_a-\delta_a)^2/\Delta ^2}+\frac{1}{2\Delta \sqrt{\pi}}e^{-(\omega-\omega_a+\delta_a)^2/\Delta ^2}$. Each Gaussian subensemble is one Zeeman branch and consists of half of the total population due to a thermal distribution at 3.6 K. The on resonance ($\delta=0$) cavity transmission spectrum can be obtained by evaluating Eq.~(3), which updates Eq.~(7) to

\begin{equation}
t(\omega) = \frac{- \frac{\kappa}{2 i}}{\omega_a - \frac{ i \kappa}{2} - \omega - i\frac{\Omega^2}{2\Delta}\sqrt{\pi}\sum_{j=1,2}e^{-(\frac{\omega-\omega_{aj}+i\gamma_h/2}{\Delta})^2}\rm erfc(-i\frac{\omega-\omega_{aj}+i\gamma_h/2}{\Delta})}.
\end{equation}
 
\noindent where $\omega_{a1}=\omega_a-\delta_a$, $\omega_{a2}=\omega_a+\delta_a$ are the central frequencies of two Zeeman branches, respectively.  

From the absorption spectrum in Fig.~1{\bf b} for 0.1\% Nd:YVO, we extract the parameters $\delta_a=$8.5 GHz, and $\Delta=$5.0 GHz. With $\kappa$ and $\gamma_h$ measured above, we plot in Fig.~2{\bf e} and Fig.~S1 the theoretical transmission (red curve) based on Eq.~(8). Note that this theoretical curve is not a fit, which clearly reveals the middle peak. The measured transmission spectrum is overlaid with the theory curve, showing good agreement.

In Fig.~2f of the main text, the polariton decay $\Gamma$ was compared to the expected decay from a Gaussian ensemble that has the same FWHM as the sum of the two Zeeman branches combined. Here we justify that this approximation of two Zeeman sub-ensembles by one broader Gaussian distribution is valid.  As mentioned in the main text, to treat the entire ensemble as one Gaussian distribution, we find the effective FWHM to be 24 GHz, and $\Delta/2\pi$=14.6 GHz. We plot the expected on resonance transmission using Eq.~(7) in blue in Fig.~S1. We see a high degree of agreement between the red and blue curves except for the middle peak region. This means the one-Gaussian approximation captures all the essential properties of the polariton spectrum. This is expected from the conclusions drawn from \cite{Diniz}, that the polariton linewidths only depend on the profile of the tails of the distribution, but not the central region of the distribution function $\rho(\omega)$. Thus we confirm the validity of the one-Gaussian approximation.
\begin{figure}[htb]
\includegraphics[width=.4\textwidth]{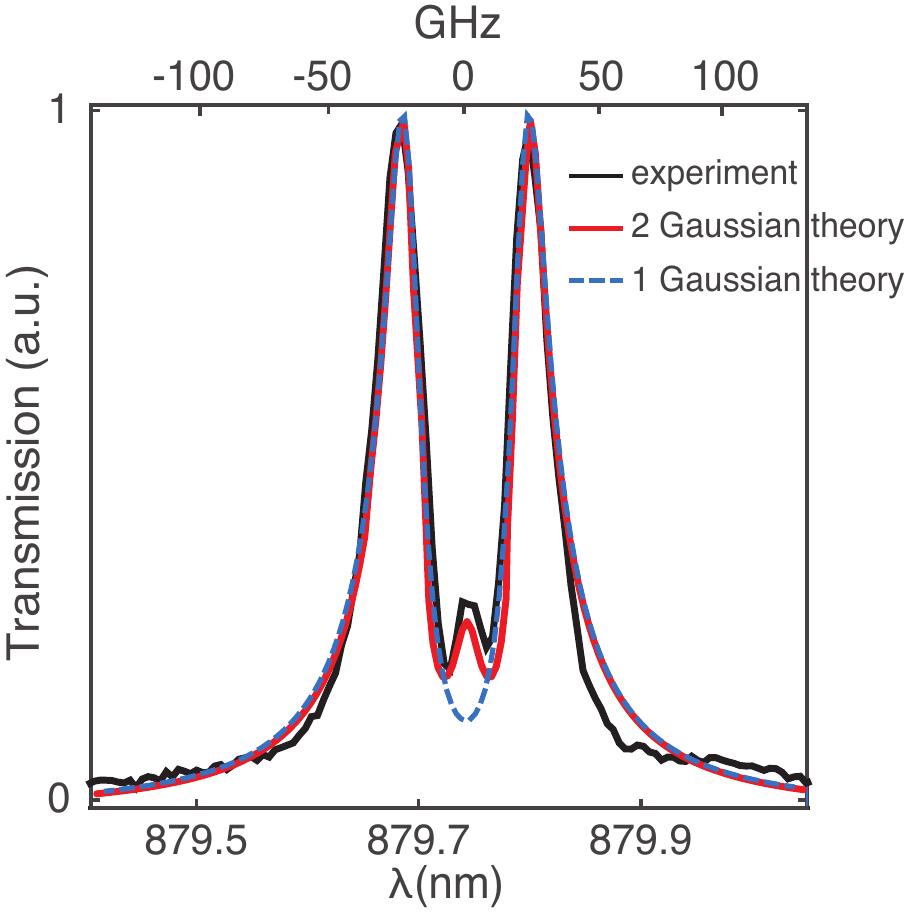}
\caption{{\bf Experimental and theoretical on-resonance spectrum in a 0.1\% Nd:YVO cavity} Black curve is the experimental spectrum. Red curve is the theoretical spectrum using a model based on two distinct Gaussian sub-ensembles, which reveals a middle peak that is consistent with the experiment. Blue dotted curve is the theoretical spectrum using a model based on one Gaussian distribution whose inhomogeneous width approximate the total width of the two Gaussian sub-ensembles. The blue curve does not show the middle peak, but otherwise coincides with the red curve well.}\label{fs1}
\end{figure}

\subsection{Experimental polariton linewidths versus detuning}
\noindent Here we plot the linewidths of both upper (shorter wavelength) and lower (longer wavelength) polaritons as the cavity was tuned from shorter to longer wavelengths across the atomic transition. For 1\% Nd:YVO, the evolution of each polariton linewidth with detuning is asymmetric about the ensemble center frequency (Fig.~S2a). Two peaks at 50 GHz, and -160 GHz detuning are evident. We believe these two peaks are due to cavity coupling to two satellite lines of main transition. Given the high density of this sample, we expect that these satellite lines correspond to the Nd-Nd pair site (resulting from Dzyaloshisky-Moriya interactions), which were measured to be +48 GHz and -166 GHz detuned from the line center~\cite{DM}. On the other hand, in 0.1\% Nd:YVO (Fig.~S2b), the linewidths of upper and lower polaritons as a function of detuning appear to be symmetric. This is expected from the symmetric shape of the ensemble distribution (Fig. 1b in the main text), and the absence of strong pair-site satellite lines at this lower doping concentration. Again, we plot in dark green the theoretical linewidths of the polaritons if the ensemble distribution were Lorentzian. At zero detuning, the linewidths of both polaritons should converge to a value $\kappa/2+\gamma_h/2+\Delta$ (i.e. the Lorentzian limit with no protection effect). In Fig.~S2a, b, the gap between the experimental zero-detuning (on-resonance) linewidths and the Lorentzian limit represents the amount of linewidth narrowing and the extent of cavity protection. This narrowing is highlighted by an red arrow.

\begin{figure}[htb]
\includegraphics[width=.8\textwidth]{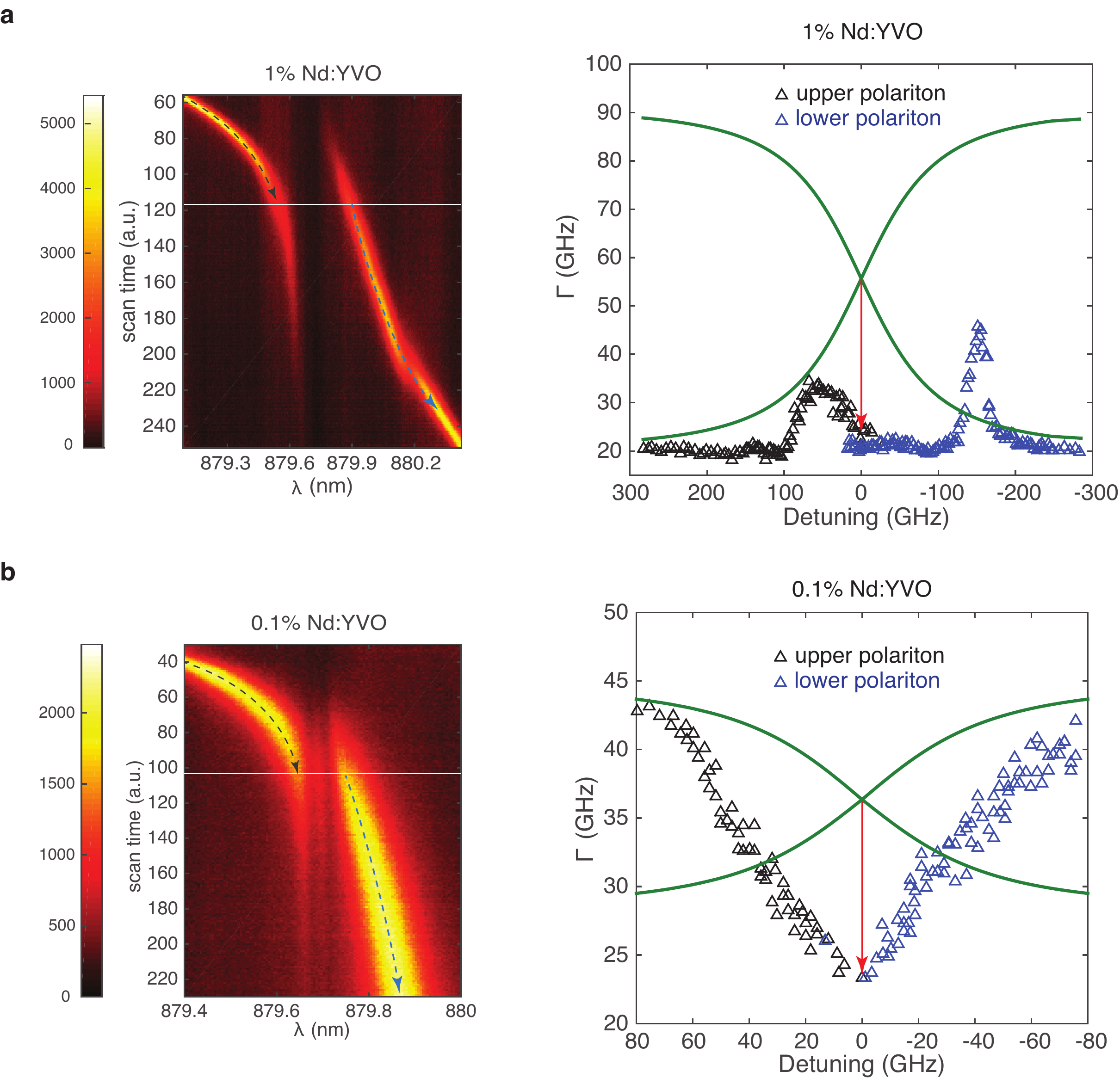}
\caption{{\bf Experimental polariton linewidths versus detuning.} {\bf a,} 1\% Nd:YVO. {\bf b,} 0.1\% Nd:YVO. Black(blue) triangles are for lower(upper) polariton. Green curves are theoretical linewidths for both polaritons assuming a Lorentzian ensemble distribution.}\label{fs3}
\end{figure}

\subsection{Simulation of the dynamics of the coupled cavity-ensemble system}
\noindent To simulate the temporal dynamics of the system (Fig.~3 theoretical plots and Fig.~4b), we start from the discrete form of the differential equations Eq.~(1). In this case, simulating the entire system of $N$ emitters entails solving a set of $N\sim10^6$ coupled differential equations. To make the problem more computationally tractable, we instead solve the system of $N_{sim}\ll N$ coupled emitters with their frequencies randomly assigned according to the experimentally measured atomic distribution $\rho(\omega)$. The coupling strength of each emitter is set as $g=\Omega/\sqrt{N_{sim}}$ such that the collective coupling strength is held constant at the experimental value $\Omega=$25 GHz. This reduced set of equations was then solved numerically in Mathematica using the built-in differential equation solver (NDSolve). The number of simulated emitters was increased until the solution converged (i.e. until increasing the number of simulated emitters no longer had an effect on the solution). 
% solve 
All emitters were assumed to start in the ground state in an empty cavity. The cavity and probe were assumed to be on resonance with the center of the emitter spectral distribution. The input field consisted of two 4(1.5) ps pulses for 0.1\% (1\%) doped cavity separated by variable time $\tau$. The amplitude of the integrated interference (corresponding to the measurement) for each value of $\tau$ was determined by integrating the cavity transmission over the simulation time (10 cavity lifetimes). 

\subsection{Relationship between the decay time constants of the interferometric signal and the polariton linewidth}
\noindent The experiments we performed to obtain Fig.~3 are optical field autocorrelation measurements at the single photon level. Here we clarify how the decay times of the interferometric signals (Fig.~3 right panel) are related to the actual polariton decay times. Given a polariton linewidth $\Gamma$, the $1/e$ intensity decay constant is $1/\Gamma$~\cite{Noda}. The field decay constant would be twice as long $2/\Gamma$. The interference of two identical but time-delayed field generates an autocorrelation function of the field. In our case, the field is exponential decaying in time. The autocorrelation of an exponential decay function is another exponential decay with twice the decay constant. Therefore, the interference signal should yield a decay constant $4/\Gamma$. We verify this relationship experimentally by measuring the decay time of an empty (far-detuned) cavity. For 0.1\% Nd:YVO cavity, $\kappa=2\pi\times$44 GHz, thus the cavity lifetime is $1/\kappa=$3.6 ps. By using the optical autocorrelation technique, we measured a decay constant of interference signal 14.5 ps, as shown in blue-dotted fit in Fig.~3{\bf a}, which is equal to $4/\kappa$.

\subsection{Temporal measurements of the cavity transmission for 1\% Nd:YVO device}
\noindent Figure S3 plots the theoretical interference fringe amplitudes along with the measured results. The mean photon number per pulse coupled in the cavity was estimated at $\mu$=0.5. The case of an uncoupled cavity is plotted in Fig.~S3a, showing a fitted decay constant ($4/\kappa$) of $\sim$31.8 ps. When only one polariton ($|0\rangle$ or $|1\rangle$) was excited, the decay was 29.8 ps (Fig.~S3b). For the superposition of two polaritons, Ramsey-like fringes were obtained, revealing at least 5 Rabi oscillations  (Fig.~S3c). The dotted blue and red curves are single exponential fit to the decaying amplitude of the fringe signal.
\begin{figure}[htb]
\includegraphics[width=.85\textwidth]{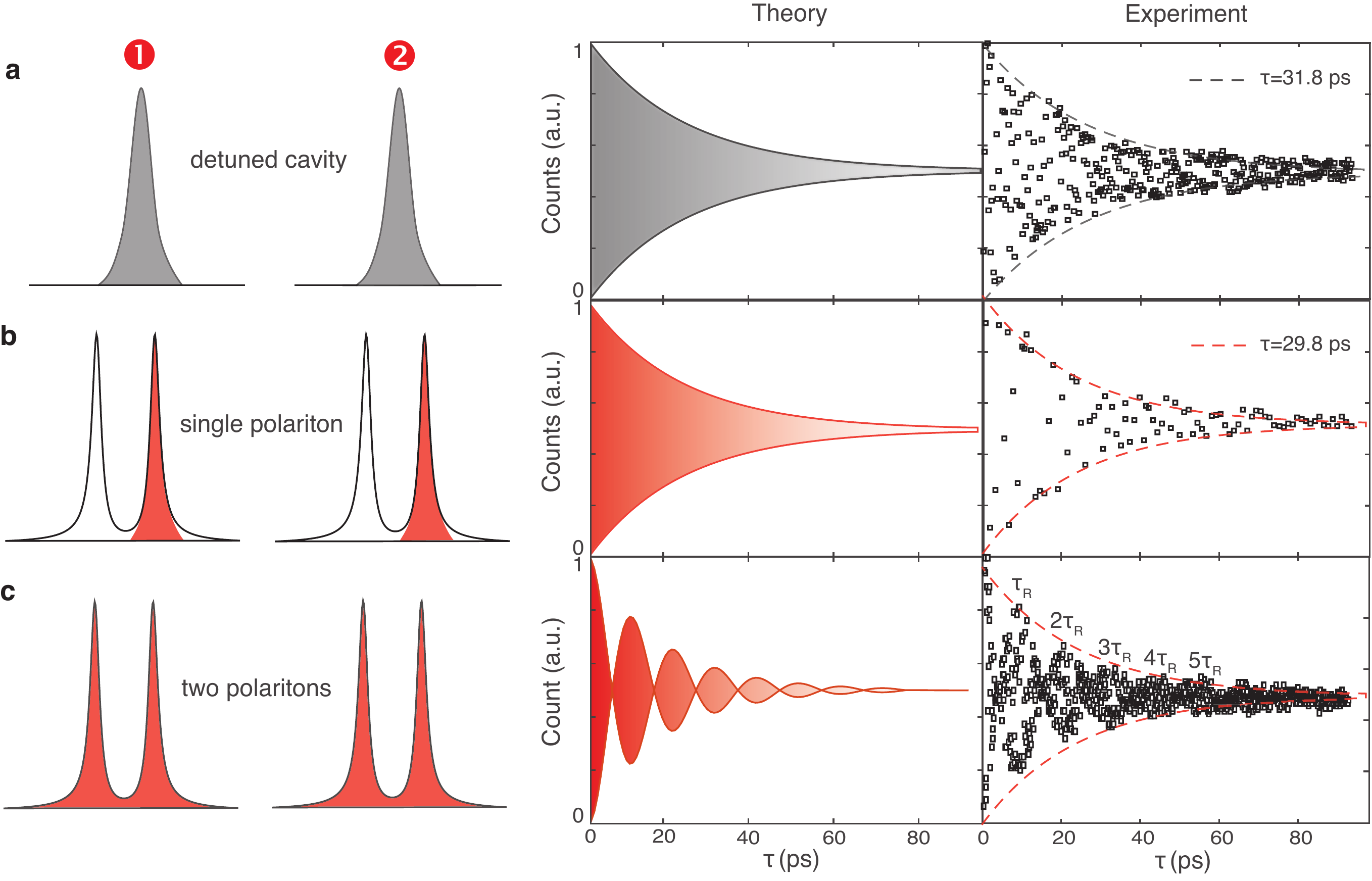}
\caption{{\bf Time-domain interferometric measurements of the cavity transmission.} Left panels show the cavity transmission spectra for the two pulses under different excitation schemes. Blue area is for uncoupled cavity; red for excited polariton states. The colored areas were plotted against the polariton spectrum (solid curve). {\bf a,} Simulated and measured cavity decay (i.e. lifetime) when uncoupled from the atoms. {\bf b,} Decay of singly excited lower (upper) polarition. {\bf c,} Decayed oscillations when both polaritons were excited with transform-limited pulses. The dotted blue and red curves are single exponential fit to the decaying amplitude of the fringe signal, from which the decay time constants are extracted. }\label{fs2}
\end{figure}

\subsection{Construction of qubit density matrix from interference fringes of two photons}
\noindent The state of an arbitrary qubit $|\psi\rangle=\alpha|0\rangle+\beta|1\rangle$, where $|\alpha|^2+|\beta|^2=1$, can be determined by taking a set of four projection measurements represented by the operators $\hat{\mu_0}=|0\rangle\langle0|+|1\rangle\langle1|$, $\hat{\mu_1}=|+\rangle\langle+|$, $\hat{\mu_2}=|\circlearrowleft\rangle\langle\circlearrowleft|$, $\hat{\mu_3}=|0\rangle\langle0|$ \cite{James}. The outcome of these measurements are
\begin{equation}
n_j=C \mathrm{Tr}\{\hat{\rho}\hat{\mu_j}\},
\end{equation}
where $\rho=|\phi\rangle\langle\phi|$ for a pure state and the scaling factor $C$ is the number of detected photons, which will be set to 1 in the following derivation. We explicitly write out the four $n_j$ values for an arbitrary qubit as
\begin{eqnarray}
n_0 &=& 0.5(| \alpha |^2+|\beta |^2)=0.5\nonumber\\
n_1 &=& 0.5|\alpha+\beta |^2\nonumber\\
n_2 &=& 0.5|\alpha-i\beta |^2\nonumber\\
n_3 &=& |\alpha |^2.
\end{eqnarray}
From these values, the four Stokes parameters are calculated as
\begin{eqnarray}
S_0&=&2n_0\nonumber\\
S_1&=&2(n_1-n_0\nonumber)\\
S_2&=&2(n_2-n_0)\nonumber\\
S_3&=&2(n_3-n_0).
\end{eqnarray}
The density matrix $\hat{\rho}$ is then constructed from the Stokes parameters by
\begin{equation}
\hat{\rho}=\frac{1}{2}\sum_{j=0}^3\frac{S_j}{S_0}\hat{\sigma}_j,
\end{equation}
where $\hat{\sigma}_0$ is the identity operator $\hat{I}$ and $\hat{\sigma}_{1,2,3}$ are the Pauli spin operators. 

Now we turn to the measurement of interference between two photonic states represented by the electric field operators $\hat{E}_1(t)=\alpha_1 e^{-i\omega_-t}a_-+\beta_1 e^{-i\omega_+t}a_+$ and $\hat{E}_2(t)=\alpha_2 e^{-i\omega_-t}a_-+\beta_2 e^{-i\omega_+t}a_+$, where $\omega_-$, $\omega_+$ are the optical frequencies of the lower and upper polaritons, respectively, with frequency difference $\omega_+-\omega_-=\Omega_R$. For simplicity, we do not include the finite linewidth of each polariton as it does not affect the results of the tomography measurement. The field operator at the single photon detector is $\hat{E}(t)=\hat{E}_1(t-\tau_R)+\hat{E}_2(t)$ corresponding to the first photon delayed by $\tau_R$ after storage in the ensemble. The count rate on the detector is
\begin{eqnarray}
C &\propto& \langle E(t)^{\dagger} E(t)\rangle \nonumber\\
&=& 2+2\cos{\phi}|\alpha_1\alpha_2^*+\beta_1\beta_2^*e^{-i\Omega_R\tau_R}|\nonumber\\
&=& 2+2 \cos{\phi}|\alpha_1\alpha_2^*+\beta_1\beta_2^*|,
\end{eqnarray} 
where the last equality holds for $\Omega_R\tau_R=2\pi$. Here $\phi$ is the carrier phase difference between the two photons, which is varied from $0$ to $2\pi$ to produce interference fringes with a peak-to-peak amplitude $C_{\rm max}-C_{\rm min}=4|\alpha_1\alpha_2^*+\beta_1\beta_2^*|$. We define a set of experimentally measurable fringe amplitude parameters $A=C_0|\alpha_1\alpha_2^*+\beta_1\beta_2^*|$, where $C_0$ is a constant factor representing the integrated counts. These parameters closely resemble the projection measurement outcomes in Eq.~(10) (different by a power of 2) depending on the states encoded on the second photon, i.e. $\alpha_2|0\rangle+\beta_2|1\rangle$.  For instance, if we encode the second photon in the same qubit as the first photon but attenuate the intensity by a factor of 2 , i.e. $\alpha_2=\alpha_1/\sqrt{2}$, $\beta_2=\beta_1/\sqrt{2}$, we get the amplitude $A_0=\frac{C_0}{\sqrt{2}}(|\alpha_1|^2+|\beta_1|^2)$. For $\alpha_2=\beta_2=1/\sqrt{2}$, $A_1=C_0/\sqrt{2}|\alpha_1+\beta_1|$. For $\alpha_2=-i\beta_2=1/\sqrt{2}$, $A_2=C_0/\sqrt{2}|\alpha_1-i\beta_1|$. For $\alpha_2=1$, $\beta_2=0$, $A_3=C_0|\alpha_1|$. Based on the four amplitude values, we construct an equivalent set of Stokes parameter $S_j^A$
\begin{eqnarray}
S_0^A &=& 2A_0^2\nonumber\\
S_1^A &=& 2A_1^2-2A_0^2\nonumber\\
S_2^A &=& 2A_2^2-2A_0^2\nonumber\\
S_3^A &=& 2A_3^2-2A_0^2,
\end{eqnarray}
from which the density matrix is calculated by
\begin{eqnarray}
\hat{\rho}&=&\frac{1}{2}\sum_{j=0}^3\frac{S_j^A}{S_0^A}\hat{\sigma_j}\nonumber\\
&=& \frac{1}{2}[\hat{I}+((A_1/A_0)^2-1)\hat{\sigma}_1+((A_2/A_0)^2-1)\hat{\sigma}_2+((A_3/A_0)^2-1)\hat{\sigma}_3].
\end{eqnarray}
Then we perform a maximal likelyhood estimation \cite{James}  to obtain a physical density matrix, which is used to calculate the fidelity $F=\langle\psi_{in}|\hat{\rho}|\psi_{in}\rangle$

\subsection{Classical storage fidelity for weak coherent photons}
\noindent The classical fidelity for any storage device measures the best input/output fidelity one can achieve using a classical method. For a given photon number of the input state $N_{ph}$, the maximum classical fidelity is known to be $F=\frac{N_{ph}+1}{N_{ph}+2}$ \cite{Massar}. For an input pulse that is in a coherent state with a mean photon number $\mu$, the Poissonian statistics give a $N$-photon probablity of $P(N_{ph})=e^{-\mu}\frac{\mu^{N_{ph}}}{N_{ph}!}$. Accounting for each $N$-photon component, the classical fidelity of a coherent state then is
\begin{equation}
F=\sum_{N_{ph}\geq1}^{\infty} \frac{N_{ph}+1}{N_{ph}+2}\frac{P(N_{ph})}{1-P(0)}
\end{equation}
In addition, for an imperfect memory with storage and retrieval efficiency $\eta<1$, the classical fidelity would be higher because a classical memory can preferentially measure the higher photon component of the input and send out a new qubit. We follow the strategy in \cite{Gundogan, Specht} that there exists a threshold photon number $N_{\rm min}$ that the classical memory sends out a qubit when the input photon number is greater than this value, which happens with a probability $1-p$. Otherwise the memory sends out a result for input photon $N_{\rm min}$ with probability $p$. Combing the two cases, the memory efficiency is expressed as
\begin{equation}
\eta=\frac{p+\sum_{N_{ph}\geq N_{\rm min+1}}P(N_{ph})}{1-P(0)}
\end{equation}
For a given $\mu$ and $\eta$, the value of $N_{\rm min}$ can be readily calculated according to \cite{Gundogan},
\begin{equation}
N_{\rm min}={\rm min}\ i: \sum_{N_{ph}\geq i+1} P(N_{ph})\leq(1-P(0))\eta,
\end{equation}
which is used to obtain the final classical fidelity
\begin{equation}
F_{\rm class}=\frac{\frac{N_{\rm min}+1}{N_{\rm min}+2}p+\sum_{N_{ph}\geq N_{\rm min+1}}\frac{N_{ph}+1}{N_{ph}+2}P(N_{ph})}{\eta(1-P(0))}
\end{equation}

\end{document}